\documentclass[twocolumn]{aastex63}

\usepackage{amsmath,amssymb,lineno}
\usepackage{newtxmath}
\usepackage{comment}

\shorttitle{Observational constraints on cosmic-ray escape from UHE accelerators}
\shortauthors{Luce et al.}

\newcommand{\dif}{\mathrm{d}}

\begin{document}

%\title{Constraints on cosmic-ray accelerators shaping the ankle feature in the energy spectrum}
\title{Observational constraints on cosmic-ray escape from ultra-high energy accelerators}

\correspondingauthor{Quentin Luce, Sullivan Marafico}
\email{quentin.luce@kit.edu, sullivan.marafico@ijclab.in2p3.fr}

\author[0000-0003-3745-6709]{Quentin Luce}
\affiliation{Karlsruhe Institute of Technology, Institute for Experimental Particle Physics (ETP),
Karlsruhe, Germany}

\author[0000-0003-4604-4677]{Sullivan Marafico}
\affiliation{Laboratoire de Physique des 2 Infinis Ir\`ene Joliot-Curie,
CNRS/IN2P3, Universit\'{e} Paris-Saclay, Orsay, France}

\author[0000-0002-4202-8939]{Jonathan Biteau}
\affiliation{Laboratoire de Physique des 2 Infinis Ir\`ene Joliot-Curie,
CNRS/IN2P3, Universit\'{e} Paris-Saclay, Orsay, France}

\author[0000-0001-5681-0086]{Antonio Condorelli}
\affiliation{Laboratoire de Physique des 2 Infinis Ir\`ene Joliot-Curie,
CNRS/IN2P3, Universit\'{e} Paris-Saclay, Orsay, France}

\author[0000-0001-6863-6572]{Olivier Deligny}
\affiliation{Laboratoire de Physique des 2 Infinis Ir\`ene Joliot-Curie,
CNRS/IN2P3, Universit\'{e} Paris-Saclay, Orsay, France}

\submitjournal{ApJ}
\received{}
\revised{}
\accepted{}

\begin{abstract}

%Interactions of ultra-high energy cosmic rays (UHECRs) accelerated in specific astrophysical environments have been shown in several studies to shape the energy production rate of injected nuclei differently from that of the secondary neutrons escaping from the confinement zone. Here, we exploit the energy spectrum and mass composition of UHECRs inferred at the Pierre Auger Observatory to constrain a benchmark scenario of in-source interactions. We obtain that the data can be  reproduced using a spatial distribution of sources that follows the density of stellar mass, providing hence a robust set of  constraints for the emission mechanisms at play in extragalactic accelerators as well as for their energetics and for the abundances of elements in their environments. While the quasi mono-elemental increase of the cosmic-ray mass number observed on Earth from ${\simeq}\: 2$\:EeV up to the highest energies calls for nuclei accelerated with a hard spectral index, the inferred flux of protons down to ${\simeq}\: 0.6$\:EeV is shown to require for this population a spectral index significantly softer than that generally obtained up to now.  

Interactions of ultra-high energy cosmic rays (UHECRs) accelerated in specific astrophysical environments have been shown to shape the energy production rate of nuclei differently from that of the secondary neutrons escaping from the confinement zone. Here, we aim at testing a generic scenario of in-source interactions through a phenomenological modeling of the flux and composition of UHECRs. We fit a model in which nucleons and nuclei follow different particle energy distributions to the all-particle energy spectrum, proton spectrum below the ankle energy and distributions of maximum shower depths above this energy, as inferred at the Pierre Auger Observatory. We obtain that the data can be reproduced using a spatial distribution of sources that follows the density of extragalactic matter on both local and large scales, providing hence a realistic set of constraints for the emission mechanisms in cosmic accelerators, for their energetics and for the abundances of elements at escape from their environments. While the quasi mono-elemental increase of the cosmic-ray mass number observed on Earth from ${\simeq}\: 2$\:EeV up to the highest energies calls for nuclei accelerated with a hard spectral index, the inferred flux of protons down to ${\simeq}\: 0.6$\:EeV is shown to require for this population a spectral index significantly softer than that generally obtained up to now. We demonstrate that modeling UHECR data across the ankle substantiate the conjecture of in-source interactions in a robust statistical framework, although pushing the  mechanism to the extreme.

\end{abstract}


\section{Introduction} 
\label{sec:intro}

The sources of ultra-high energy cosmic rays (UHECRs) remain unknown. Their identification relies primarily on capturing in the UHECR arrival directions a pattern suggestive in an evident way of a class of astrophysical objects. Such a capture is still eluding our grasp, but some recent observations have allowed broad statements to be drawn. An anisotropy at large scales has been revealed above ${\simeq}\:8\:$EeV~\citep{PierreAuger:2017pzq}, the contrast and the direction of which are consistent with expectations drawn from sources distributed in a similar manner to the extragalactic matter~\citep{PierreAuger:2017pzq,PierreAuger:2018zqu}. At higher energies (${\gtrsim}\:40\:$EeV), a correlation between UHECR arrival directions and the flux patterns of massive, star-forming or active galaxies within 200\:Mpc provide evidence for anisotropy~\citep{PierreAuger:2018qvk,PierreAuger:2021rfz}. Overall, these observations suggest that UHECRs are predominantly of extragalactic origin at least above the so-called ankle energy at ${\simeq}\:5\:$EeV. 

The energy spectrum and chemical composition of UHECRs observed on Earth result from the emission processes at play, which encompass acceleration mechanisms, losses and escape from the source environments as well as propagation effects. Differently from, and complementary to anisotropies, these two observables provide constraints helping infer the properties of the acceleration processes, the energetics of the sources, and the abundances of elements in the source environments. Following this principle, several studies have laid the foundations for a generic scenario that broadly reproduces the observations~\citep{Allard:2008gj,Aloisio:2013hya,Taylor:2015rla,PierreAuger:2016use,Zhang:2017moz,PierreAuger:2021mmt}. The intensity of the individual nuclear components at the sources, generally considered as stationary and uniformly distributed in a comoving volume, is assumed to drop off at the same magnetic rigidity so as to explain the gradual increase with energy of mass number $A$ observed on Earth~\citep{PierreAuger:2010ymv,PierreAuger:2014gko,Watson:2021rfb}. This is consistent with the basic expectation that electromagnetic processes accelerate particles up to a maximum energy proportional to their electric charge $Z$. Most notably, the abundance of nuclear elements is found to be dominated by intermediate-mass ones, ranging from He to Si, accelerated to $E^{Z}_{\mathrm{max}}\simeq 5Z\,\text{EeV}$ and escaping from the source environments with a very hard spectral index $\gamma$ (that characterizes the emission spectrum at the sources as $E^{-\gamma}$), which, depending on the systematic uncertainties affecting some of the necessary modeling, ranges between $\gamma \simeq -1$ and $\gamma \simeq 1$~\citep{PierreAuger:2016use}. 

Reaching ultra-high energies can be most easily achieved by first-order Fermi shock acceleration, where particles are energized by bouncing back and forth across moving magnetic fields~\citep[see e.g.][for a review]{Sironi:2015oza}. While the typical spectral index for this mechanism ($\gamma \gtrsim 2$) appears to deviate significantly from the hard values favored from the data, the interplay between the escape from the sources and the acceleration mechanisms must be taken into account to correctly apprehend the emission spectrum~\citep[e.g.][]{Fujita:2009ak}, in particular when escape is influenced by in-source interactions~\citep{Unger:2015laa,Globus:2015xga,Biehl:2017zlw,Zhang:2017moz,Fang:2017zjf,Supanitsky:2018jje,Boncioli:2018lrv,Muzio:2021zud}. The cosmic-ray luminosity of the source candidates is governed, among other things, by the levels of radiation and magnetic-field densities. While gaining energy, UHECRs can interact with this radiation or escape from the magnetized zone. Thus, the energy spectrum of the ejected particles as well as the amount of ejected nuclei may differ strongly from those injected into the electromagnetic field. This has two important consequences: a) the ejected spectrum of the charged nuclei can be much harder than that injected, due to the escape mechanism and in particular the behavior of the nuclei-photon cross section at high energies; b) the interactions can produce a copious flux of secondary neutrons of energy $E_n = E/A$. These neutrons can escape freely from the magnetic confinement zones, with an ejection spectrum much softer than that of nuclei, to decay into protons on their way to the Earth after an average travel distance of $9.1\:\text{kpc}\times(E_n / 1\:\text{EeV})$. 

In this paper, we derive constraints on the characteristics of the sources by searching for a robust statistical agreement between data and predictions based on realistic astrophysical ingredients and on a phenomelogical model incorporating in-source interactions in an effective way. The agreement is sought for both spectral and mass composition data, while only a qualitative overview of the latter has been performed so far in studies incorporating in-source interactions. To do this, we consider the data obtained at the Pierre Auger Observatory beyond 0.63\:EeV ($10^{17.8}~$eV). However, and differently from the approaches adopted elsewhere, only the proton spectrum is used in the energy range between 0.63\:EeV and 5~EeV. This approach allows us to reconstruct the extragalactic component without resorting to any introduction of \textit{ad hoc} nuclear components to model what is generally assumed to be the upper end of the galactic component so as to reproduce the average mass composition as a function of energy. Such a modeling, suffering from both observational systematics and theoretical unknowns, blurs the reconstruction of the extragalactic proton component with biases inherent to the choices made. In the same spirit, we do not attempt to sew together the Auger observations with those made in an energy range half-a-decade lower with the KASCADE-Grande detector mixing indistinctly protons and Helium~\citep{Apel:2012tda}, always with the aim of reconstructing the proton component as precisely as possible. 

With this approach, we show that a scenario incorporating in-source interactions can reproduce the data and that accounting for the local overdensity of galaxies within tens of Mpc further improves the goodness of fit with respect to the generally assumed uniform distribution of sources in a comoving volume. Quantitatively, 
we show that the analysis of data from the Pierre Auger Observatory significantly favors an effective spectral index of protons ejected from the sources that is softer than generally obtained or assumed, somehow pushing the in-source interaction mechanism to the extreme. At the same time, we point out that this spectral index approaches more ``canonical'' values provided that all protons are secondaries of the interactions, i.e.\ that the environment of the sources is devoid of protons to be accelerated. 

This paper is organized as follows. In Section~\ref{sec:model}, we motivate the model used to probe those effects and set up a benchmark scenario among the ingredients needed to propagate the particles from their sources to the Earth. The data and fitting method are presented in Section~\ref{sec:data}. Results obtained with the benchmark scenario are given in Section~\ref{sec:results}, where we explore as well the impact of changing the ingredients that are affected by systematic uncertainties. Finally, the significance of the results is discussed in Section~\ref{sec:discussion}.

\section{Generic model of UHECR production with in-source interactions} 
\label{sec:model}

The benchmark astrophysical model used here is inspired by that of~\cite{PierreAuger:2016use} for its main features. The non-thermal processes responsible for accelerating the different particles are modeled through power-law spectra as long as energies are sufficiently below the $Z$-dependent maximal acceleration energy $E^{Z}_{\mathrm{max}}=ZE_{\mathrm{max}}$, while, in the absence of any firmly-established prescription from theory, an exponential suppression is used for modeling the upper end of the acceleration process. The sources are assumed to accelerate different amounts of nuclei represented by five stable ones: hydrogen ($^1$H), helium ($^4$He), nitrogen ($^{14}$N), silicon ($^{28}$Si) and iron ($^{56}$Fe). The unavoidable fluctuations of the characteristics of each individual source are neglected, considering all sources as identical. This is a simplification that renders the number of free parameters manageable in a fitting procedure. The fitted parameters should be interpreted as effective ones, describing the average spectrum of the source population.

Modeling the ejection spectra in terms of power laws suppressed at the highest energies is an approximation aimed at keeping, once again, the number of free parameters to a minimum. However, detailed studies of the nuclear cascade developing in sources from photodissociation with different levels of radiation densities, such as that presented in~\cite{Biehl:2017zlw}, show that such a simplification holds in the energy range of interest for this study. This is also true for the softer flux of secondaries neutrons produced during the development of the cascade, which dominates over that of ejected protons. We generically model the ejection rate per comoving volume unit and per energy unit of nucleons as
\begin{equation}
\label{eqn:qN}
    q_{\rm p}(E) = q_{0{\rm p}}\left(\frac{E}{E_0}\right)^{-\gamma_{\rm p}}f_{\mathrm{supp}}(E,Z_{\rm p}),
\end{equation}
with $Z_{\rm p}=1$, that is a single reference ejection rate $q_{0{\rm p}}$, spectral index $\gamma_{\rm p}$ and suppression function $f_{\mathrm{supp}}(E,Z_{\rm p})$ for both escaping protons and protons from neutron decay. Here, $E_0$ is arbitrarily set to 1\:EeV. The ejection rate of nuclei with mass number ${A}_i$ is also generically modeled as
\begin{equation}
\label{eqn:qA}
    q_{{A}_i}(E) = q_{0{A}_i}\left(\frac{E}{E_0}\right)^{-\gamma_{A}}f_{\mathrm{supp}}(E,Z_{{A}_i}),
\end{equation}
that is a single spectral index $\gamma_{A}$ and four independent reference ejection rates $q_{0{A}_i}$ for helium, nitrogen, silicon and iron. As in the reference scenario of \cite{PierreAuger:2016use}, the suppression function adopted both for nucleons and nuclei is taken as
\begin{equation}
\label{eqn:fsupp}
    f_{\mathrm{supp}}(E,Z) = 
    \begin{cases}
    1 & \mathrm{if~}E\leq E^{Z}_{\mathrm{max}},\\
    \exp{\left(1-E/E^{Z}_{\mathrm{max}}\right)} & \mathrm{otherwise}.
    \end{cases}
\end{equation}
The maximum acceleration energy is assumed to be proportional to the electric charge of each element, $E^{Z}_{\mathrm{max}} = ZE_{\mathrm{max}}$, with a single free parameter $E_{\mathrm{max}}$ shared by the five species.

In this approach, the ejection rate for protons accounts for both the accelerated ones up to $E_{\mathrm{max}}$ and those produced by the escaping neutrons, inheriting hence an energy of the order of $E/A$ from the nuclei of energy $E$. The maximum energy that this population of secondary protons can reach is then $E^{Z}_{\mathrm{max}}/A\simeq E_{\mathrm{max}}/2$, which is not reflected in the rigidity-acceleration scheme modeled by Equation~\eqref{eqn:fsupp}. The extreme case in which all ejected protons would be photodissociation by-products can be tested by replacing $E_{\mathrm{max}}$ by $E_{\mathrm{max}}/2$ in $q_{\mathrm{p}}(E)$. We will explore this extreme case in Section~\ref{sec:results}. Further characterization of the balance between the population of accelerated protons from the initial abundance in the source environment and that of secondaries from nuclear cascades calls for modeling of specific sources and environments, which is beyond the scope of this paper. 

The differential energy production rate per comoving volume unit of the sources, which is directly related to their differential luminosity, is then $\ell_j(E,z)= E^2q_j(E)S(z)$ for each species $j$, where $S(z)$ describes the redshift evolution of the UHECR luminosity density. The quantity $\ell_j(E,z)$ is hereafter called differential energy production rate for convenience. On the other hand, the bolometric energy production rate per comoving volume unit at redshift $z$ is obtained as $\mathcal{L}_j(E, z)=S(z)\int_{E}^{\infty}\dif E'E'q_j(E')$. We hereby report its average value in a volume spanning $z_\text{min}-z_\text{max}$ as $\bar{\mathcal{L}}_j(E) = \int_{z_\text{min}}^{z_\text{max}} \dif z  \left|\frac{\dif t}{\dif z}\right| \mathcal{L}_j(E, z) / \int_{z_\text{min}}^{z_\text{max}} \dif z  \left|\frac{\dif t}{\dif z}\right|$, where $t(z)$ is the lookback time.

The evolution of the UHECR luminosity density is taken as being traced by the density of baryonic matter over cosmic time. The latter is assumed to follow the density of stellar mass, which is fairly approximated by a constant out to redshift $z=1$ \citep[e.g.][]{2014ARA&A..52..415M}. As in the benchmark scenario of \cite{PierreAuger:2016use}, such a constant evolution is assumed to hold to first approximation out to $z_\text{max} = 2.5$. The local Universe presents however an overdensity because, like most galaxies, the Milky Way belongs to a group of galaxies, itself embedded in the Local Sheet \citep{2014MNRAS.440..405M}. We adopt here the overdensity correction factor inferred by~\cite{2019ApJ...872..148C} expressed as a function of the distance $r$ and effective up to 30\:Mpc,
\begin{equation}
\label{eqn:rho}
    \frac{\delta\rho(r)}{\bar \rho} = 1+\left(\frac{r}{r_0}\right)^{-\alpha},
\end{equation}
with $r_0=5.4\:$Mpc and $\alpha=1.66$. The evolution of the UHECR luminosity density is then described as $S(z) = \delta\rho/\bar \rho$, with $z_\text{max} = 2.5$ and $z_\text{min} = 2\times 10^{-4}$.
The minimum redshift, which corresponds to the limit of the Local Group of galaxies ($r\simeq 1\:$Mpc), prevents any divergence in Equation~\eqref{eqn:rho} and effectively removes very-closeby galaxies that would otherwise dominate the UHECR sky, at odds with observations. 

Extragalactic magnetic fields are poorly known, except for upper limits at the nG level from rotation measures~\citep{Pshirkov:2015tua,2020MNRAS.495.2607O} or down to tens of pG for magnetic fields of primordial origin that would affect CMB anisotropies \citep{2019PhRvL.123b1301J}.  Lower limits at the fG level have also been derived from the non-observation in the GeV range of gamma-ray cascades from TeV blazars~\citep{Neronov2010,Tavecchio:2010mk,2018ApJS..237...32A}. These non-observations have been interpreted as the deflection of the cascade flux into a broadened beam weakening the point-like image, noting though that the lower limits could be alleviated if plasma instabilities dominated the cooling rate of particles in the cascade \citep[see][for recent reviews]{2021Univ....7..223A, 2022Galax..10...39B}. While field values saturating the limits from rotation measures might impact the results by reducing the UHECR horizon~\citep{Mollerach:2013dza,Wittkowski:2017okb}, we assume here that magnetic fields in cosmic voids are at the level of 0.1\:nG or less, as suggested by cosmological magneto-hydrodynamical simulations \citep{2017CQGra..34w4001V}. In this case, they do not have sizeable effects on the propagation of the particles, which can therefore be considered one-dimensional. 

The all-particle energy spectrum $J(E)$ observed at present time thus results from the integration of the contribution of all sources over lookback time, the role of which is played by redshift:
\begin{equation}
\label{eqn:model}
    J(E) = \frac{c}{4\pi}\sum_{A,A'} \iint \dif z\,\dif E' \left|\frac{\dif t}{\dif z}\right|S(z) q_{A'}(E')\frac{\dif\eta_{AA'}(E,E',z)}{\dif E}.
\end{equation}
Here, the relationship between cosmic time and redshift follows from the concordance model used in cosmology, $(\dif t/\dif z)^{-1}=-H_0(1+z)\sqrt{\Omega_{\mathrm{m}}(1+z)^3+\Omega_{\Lambda}}$, where $H_0=70\:$km\:s$^{-1}$\:Mpc$^{-1}$ is the Hubble constant at present time, $\Omega_{\mathrm{m}}\simeq0.3$ is the density of matter (baryonic and dark matter) and $\Omega_{\Lambda}\simeq0.7$ is the dark-energy density. The energy losses and spallation processes are described by $\eta_{AA'}(E,E',z)$, which is the fraction of particles detected on Earth with energy $E$ and mass number $A$ from parent particles emitted by the sources with energies $E'>E$ and mass numbers $A'>A$. In practice, for a given source with redshift $z_0$ emitting a nuclear species $A_0$ at energy $E_0$, the corresponding $\eta_{A_0A'}(E_0,E',z_0)$ function is tabulated in bins of $A'$ and $E'$ by propagating a large number of emitted particles (O($10^7$) particles) using the SimProp package~\citep{Aloisio:2012wj,Aloisio:2015sga,Aloisio:2016tqp}. By repeating the simulations for different values of $z_0$, $A_0$ and $E_0$, the whole $\eta_{AA'}(E,E',z)$ function is tabulated as a 5D histogram, providing hence the fractions sought for. The relevant processes accounted for in SimProp are pair production, photo-pion production and photodissociation off the photon fields of interest, which are those from the cosmic-microwave-background (CMB) radiation, and the infrared photons from the extragalactic background light (EBL) that comprises the radiation produced in the Universe since the formation of the first stars. The CMB radiation is characterized as a black-body spectrum with a redshift-dependent temperature $T(z)=T_0(1+z)$, with $T_0=2.725\:$K. The EBL is less precisely known, especially in the far infrared and at high redshifts. For the benchmark scenario explored below, we use the model of~\cite{Gilmore:2011ks} that is consistent with the minimal intensity level from galaxy counts and with indirect measurements from absorption of gamma-rays at multi-TeV energies \citep[e.g.][for a recent review]{2021arXiv211205952P}. Alternative EBL models are studied as sources of systematic uncertainties. Pair production is a very well-known process that results in a small fractional energy loss above a threshold of ${\simeq}\:0.5 A\epsilon_{\mathrm{meV}}\:$EeV, with $\epsilon_{\mathrm{meV}}$ the energy of the photon targets in meV. Photo-pion cross sections have been measured in accelerator-based experiments and are well reproduced by various event-generator codes. The corresponding fractional energy loss is quite important above a threshold of ${\simeq}\:30A\:$EeV, causing the GZK effect. Photodissociation cross sections for nuclei, on the other hand, are less known, especially for exclusive channels in which charged fragments are ejected. The fractional energy losses used here, which also shape the suppression of the spectrum, rely on phenomenological approaches, using the TALYS model~\citep{Koning:2005ezu,Koning:2012zqy} for the benchmark scenario. We also evaluate alternative models as sources of systematic uncertainties. 
%The dominant interactions are due to the giant dipolar resonance for photon energies below 30\:MeV (in the nucleus restframe) so that the center-of-mass energy is of the order of several MeV, the energy necessary to split off an individual nucleon. For photon energies between 30\:MeV and 150\:MeV, the quasi-deuteron processes cause the emission of multiple nucleons. 
In any event, because the binding energies are of order of a few MeV for all nuclei of interest, the center-of-mass energy depends only on the Lorentz factor of the nuclei. For heavy nuclei like iron, the  photodissociation thus occurs on higher-energy photons -- infrared background photons with energies about an order of magnitude higher than the typical CMB photon energy. Since these photons are less abundant than the CMB ones, the attenuation of heavy nuclei is relatively slow and is comparable to that of protons. On the other hand, the Lorentz factor is high enough for light nuclei to photodissociate on the CMB background and their attenuation is much faster than that of protons. Finally, the adiabatic losses due to the expansion of the Universe are included in the energy-loss rate of the particles as $-(1/E)(\dif E/\dif t)=-((1+z)\dif t/\dif z)^{-1}$. 

With these ingredients, Equation~\eqref{eqn:model} can be used to evaluate the all-particle flux from the various contributions of each individual nuclear component on the condition to assign values to the five ejection rates $q_{0A_i}$, the two indices $\gamma_{\rm p}$ and $\gamma_A$, and the maximum energy $E_{\rm max}$. These eight parameters are fitted to the data in the way explained in the next section.

\section{Combined fit to energy-spectrum and mass-composition data} 
\label{sec:data}

The expected spectrum modeled by Equation~\eqref{eqn:model} depends on several unknown parameters characterizing properties of the acceleration processes, of the source environments and of the source energetics. The observed energy spectrum and mass composition can, as discussed in Section~\ref{sec:intro}, provide constraints helping in inferring these parameters. 

The data we use hereafter are those obtained at the Pierre Auger Observatory, located in the province of Mendoza (Argentina) and providing the largest exposure to date to UHECRs~\citep{PierreAuger:2015eyc}. The Observatory is a hybrid system that detects the extensive air showers induced in the atmosphere subsequent to the collisions of UHECRs with nitrogen and oxygen molecules. We shall make use on the one hand of the all-particle energy spectrum inferred from these data~\citep{PierreAuger:2021hun}, and on the other hand of the distributions of the slant depth of maximum of shower development ($X_{\mathrm{max}}$), which is a proxy, the best up to date, of the primary mass of the particles~\citep{PierreAuger:2014gko,Bellido:2017cgf}. With caution over the hadronic-interaction generators  used to model the development of the showers, the $X_{\mathrm{max}}$ distributions allow for inference of the energy-dependent mass composition on a statistical basis. Two hadronic-interaction generators are considered here, namely EPOS-LHC~\citep{Pierog:2013ria} for the benchmark scenario and Sibyll2.3c~\citep{riehn2017hadronic} as an alternative, which are those up-to-date that best describe the data. Hence, the various mass components $J_j(E)$ can be derived by combining the all-particle energy spectrum and the abundances of the different elements as a function of energy. The energy threshold considered here is the nominal one of the detection mode of $X_{\mathrm{max}}$ at ultra-high energies, namely ${\simeq}\: 0.63\:$EeV ($10^{17.8}\:$eV). Incidentally, this threshold allows us to explore the energy range of the ankle feature.% by assuming no overlap for the proton component from a high-energy tail of the bulk of Galactic cosmic rays. 

From a set of proposed parameters $\mathbf{\Theta}$, the best match between the observed spectrum and that expected from Equation~\eqref{eqn:model} is obtained through a Gaussian likelihood fit. Under the assumption that the transition to extragalactic UHECRs is already completed above 5\:EeV,\footnote{We checked that a moderate increase in the value at which the transition to extragalactic UHECRs is assumed to be completed has no significant impact on the results.}  the corresponding likelihood term, $L_J$, is calculated above that threshold. The model is fitted to the $X_{\mathrm{max}}$ data, following~\cite{PierreAuger:2016use}, using the multinomial distribution that describes how likely it is to observe, in each energy bin $m$, $k_{mx}$ events out of $n_m$ with probability $p_{mx}$ in each $X_{\mathrm{max}}$ bin. The probabilities $p_{mx}$ are obtained by using generators of hadronic interactions to model the $X_{\mathrm{max}}$ distributions expressed in terms of the proposed parameters $\mathbf{\Theta}$. The corresponding likelihood term, $L_{X_{\mathrm{max}}}$, is built as the product over energy bins above 5\:EeV. The last contribution to the likelihood stems from the sub-ankle proton component, $L_{J_{\mathrm{p}}}$, obtained by weighting the all-particle spectrum below 5\:EeV with the proton abundance $f_{\mathrm{p}}(E)$ from the $X_{\mathrm{max}}$ distributions. The statistical uncertainties in $\hat{J}_{\mathrm{p}}(E)$ are dominated by those in $f_{\mathrm{p}}(E)$ %, which have no bin-to-bin correlations. Hence the likelihood is  calculated using a diagonal
and are accounted for through a Gaussian likelihood fit. 

The model likelihood is therefore given by $L=L_JL_{X_{\mathrm{max}}}L_{J_{\mathrm{p}}}$. The goodness-of-fit is assessed with a deviance, $D$, defined as the negative log-likelihood ratio of a given model and the saturated model that perfectly describes the data:
\begin{equation}
\label{eqn:deviance}
D=-2\log{\frac{L_J}{L_J^{\mathrm{sat}}}}-2\log{\frac{L_{X_{\mathrm{max}}}}{L_{X_{\mathrm{max}}}^{\mathrm{sat}}}}-2\log{\frac{L_{J_{\mathrm{p}}}}{L_{J_{\mathrm{p}}}^{\mathrm{sat}}}}.
\end{equation}
The three different contributions are referred to as $D_J$, $D_{X_{\mathrm{max}}}$ and $D_{J_{\mathrm{p}}}$ respectively.

\begin{deluxetable*}{c|cccccccc|c}[t]
\tablecaption{Best-fit parameters and minimum deviance. \label{tab:epos-results}}
\tablehead{
\colhead{Scenario} 	& \colhead{$\gamma_{\rm p}$} 	& \colhead{$\gamma_{A}$} 	&	\colhead{$\log_{10}\: E_{\rm max}\:\text{[eV]}$}	&	\colhead{$\bar{\mathcal{L}}_{\rm p}/\mathcal{L}_{0}$} 	&	\colhead{$\bar{\mathcal{L}}_{\rm He}/\mathcal{L}_{0}$} 	&	\colhead{$\bar{\mathcal{L}}_{\rm N}/\mathcal{L}_{0}$} 	&	\colhead{$\bar{\mathcal{L}}_{\rm Si}/\mathcal{L}_{0}$} 	&	\colhead{$\bar{\mathcal{L}}_{\rm Fe}/\mathcal{L}_{0}$} 	 	&	\colhead{$D/\text{ndf}$}}
\startdata
Benchmark & $3.24 \pm 0.10$    	& $-0.46\pm0.03$  	&	 $18.35\pm 0.01$  &	 $5.51 \pm 0.22$	&	 $1.71 \pm 0.14$ 	&	 $2.99\pm 0.17$ 	&	 $0.59 \pm 0.14$  	&	 $0.02 \pm 0.03$  	&	  $236.8/125$  	\\
$E_{\rm max}^{\rm p} = E_{\rm max}/2$ & $2.47 \pm 0.19$  & $-0.42 \pm 0.02$ & $ 18.37 \pm 0.01 $ & $5.23\pm0.37$ & $1.68\pm 0.14$ & $3.14\pm0.47$ & $0.50 \pm 0.99$ & $0.05\pm0.04$ & $235.8/125$\\
No overdensity &  $3.25 \pm0.14$ & $-0.68  \pm 0.03$ & $18.32 \pm 0.01$ & $5.41 \pm 0.21$ & $1.86 \pm 0.14$ & $2.72\pm 0.14$ & $ 0.95 \pm 0.25 $ & $0.09 \pm 0.14 $ & $256.9/125$ \\
$\gamma_{\rm p}=\gamma_{A}$ & $\gamma_{A}$ & $ -1.54\pm 0.10$  &	$ 18.19\pm 0.01 $  &	$ 0.63\pm0.07 $	& $ 1.44\pm0.14 $ &	 $ 3.09\pm 0.08 $ &	 $ 0.34\pm 0.21$  	& $ 0.12\pm 0.02$  & $862.7/126$\\
\enddata
\tablecomments{Column one lists the scenarios: the benchmark features different spectral indices for nucleons and nuclei, a cut-off energy for each component at $E_{\rm max}^Z = ZE_{\rm max}$ as well as an evolution of sources that is flat at large distances and follows the local overdensity of matter within 30\,Mpc. Subsequent lines provide the results for scenarios differing from the benchmark as follows: a nucleon maximum energy equal to that of escaping neutrons ($E_{\rm max}^{\rm p} = E_{\rm max}/2$), a strictly flat evolution (no overdensity) and a shared index across nucleons and nuclei ($\gamma_{\rm p}=\gamma_{A}$). Columns two and three provide the spectral indices of nucleons and nuclei, respectively. Column four provide the maximum energy of nucleons. The subsequent five columns provide the bolometric energy production rate for each species above $10^{17.8}\:$eV, normalized to a reference value $\mathcal{L}_{0} = 10^{44}\: {\rm erg}\: {\rm Mpc}^{-3}\: {\rm yr}^{-1}$. The last column provides the deviance obtained with the best-fit parameters as well as the number of degrees of freedom (ndf). The uncertainties on the parameters of the $E$- and $Z$-dependence of the ejection rate (col.\ 2--4)  are obtained through a profile likelihood. Those on the bolometric energy production rate (col.\ 5--9) are obtained from the inverse Hessian matrix, for parameters in col.\ 2--4 fixed to their best-fit values.}
\end{deluxetable*}

\section{Results} 
\label{sec:results}

Fitting the model to the data within the framework of the benchmark scenario described in Section~\ref{sec:model}, with free spectral indices for both nucleons and nuclei ($\gamma_{\rm p}\neq\gamma_{A}$), leads to the parameters and deviance given in Table~\ref{tab:epos-results} using EPOS-LHC to interpret $X_{\mathrm{max}}$ data (results obtained using Sibyll2.3c are given in Appendix~\ref{app:sibyll2.3} for completeness). For reference, Table~\ref{tab:epos-results} also provides the results obtained under the assumption of a proton component dominated by neutron escape (proton maximum energy of $E_{\rm max}/2$), of no local overdensity (widely-used uniform distribution) and of a shared spectral index across the five species ($\gamma_{\rm p}=\gamma_{A}$).

Similarly to that found in~\cite{PierreAuger:2017pzq}, the value of $E_{\mathrm{max}}$, determined by the drop in the nuclear components at $E^Z_{\mathrm{max}}$, implies that the suppression of the spectrum is due to the combination of the cut-off energy at the sources for the heavier nuclei and the energy losses \emph{en route}. The spectral index of the nuclei, $\gamma_{A}$, is in turn determined by the increase of the average mass with energy, which is almost monoelemental, so as to reproduce the $X_{\mathrm{max}}$ distributions as well as possible. The solution provided by the fit therefore consists in imposing a hard index for nuclei so that the contribution of each element mixes as little as possible: high-energy suppression imposed by the cut-off beyond $E^{Z}_{\mathrm{max}}$ and low-energy suppression via the hard index $\gamma_{A}$. However, this phenomenon does not apply to protons, which are present in an energy range where a mixture of elements is required. The best-fit value of $\gamma_{\mathrm{p}}$ is much softer than that of $\gamma_{\mathrm{A}}$. Note that the introduction of $\gamma_{\mathrm{p}}$ significantly improves the fit of the data down to 0.63\:EeV, with a total deviance $D=236.8$ compared to $862.7$ in the case of $\gamma_{\mathrm{p}}=\gamma_{\mathrm{A}}$ so that the introduction of this extra free parameter is amply justified. On the other hand, the introduction of the local overdensity to trace the source distribution also provides a substantial improvement of the deviance, although with moderate impact on the best-fit parameters. 

\begin{figure}[t]
\centering
\includegraphics[width=\columnwidth]{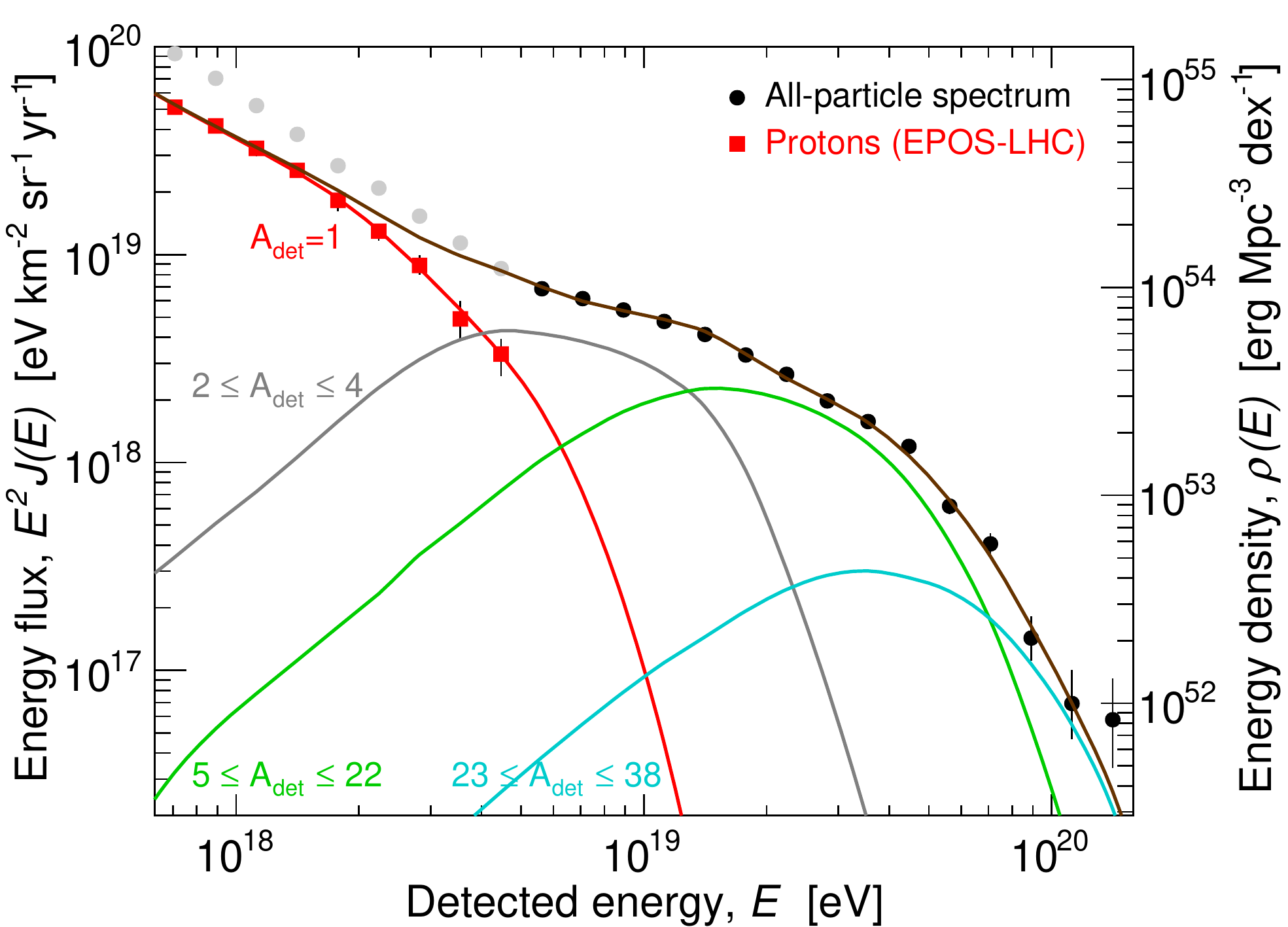}
\caption{Energy flux at Earth as a function of energy, as modeled by the best-fit parameters for the benchmark scenario. The all-particle spectrum and proton component are shown as black circles and red squares, respectively. Lighter points are not included in the fit. The best-fit components obtained for five detected mass groups are displayed with solid colored lines, as labeled in the Figure. The energy flux of the heaviest mass group, with detected mass number in 39--56, is below the range of interest.}
\label{fig:FitFlux}
\end{figure}

\begin{figure}[t]
\centering
\includegraphics[width=\columnwidth]{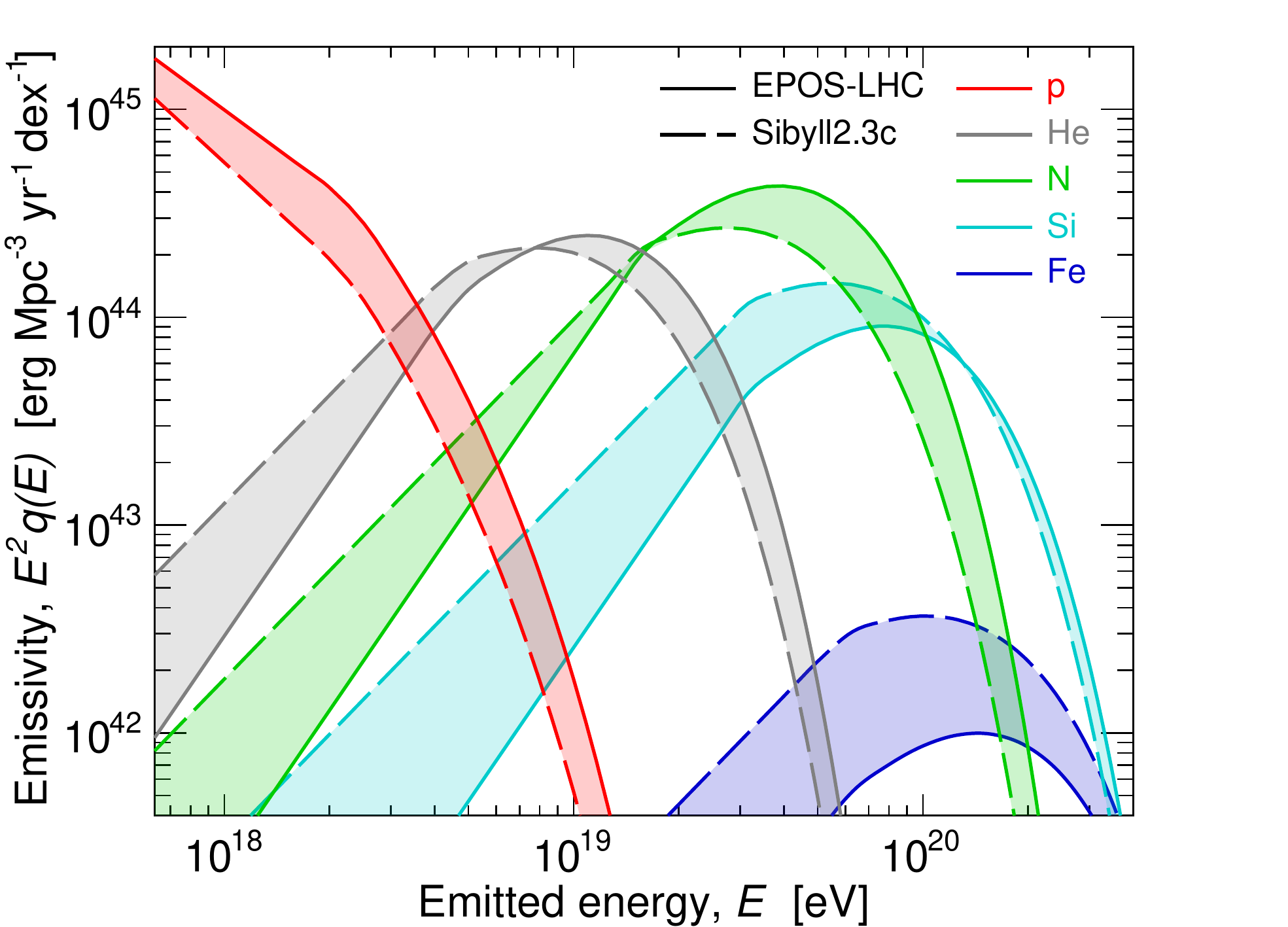}
\caption{Mass-dependent energy production rate at the sources as a function of energy, as constrained by the best-fit parameters for the benchmark scenario. The dashed lines illustrate a variation of the hadronic interaction model, with EPOS-LHC and Sibyll2.3c shown as solid and dashed lines, respectively.}
\label{fig:EProdRate}
\end{figure}

The balance regulating the intensity of each component is reported in terms of the energy production rates $\bar{\mathcal{L}}$ above 0.63\:EeV. %Note the strong correlation between these different values (see Appendix~\ref{app:tables}), correlation reflecting the intrinsic difficulty to discriminate between each intermediate mass from the $X_{\mathrm{max}}$ distributions. 
The solution is illustrated in Figure~\ref{fig:FitFlux},
%\footnote{The proton component, inferred from the mass fractions in \cite{Bellido:2017cgf}, displays highly asymmetric error bars at the highest energies. The lower statistical uncertainties on the proton flux are significantly smaller (factor of ${\sim}\:10$) than the upper uncertainties above 10\:EeV, which can be understood from the low-count regime and fractions close to zero-boundary. As shown in \cite{Bellido:2017cgf}, the proton fractions  above 10\:EeV are compatible with zero at the $1\:\sigma$ level when accounting for systematic uncertainties.}
where the contributions of each nuclear component to the observed energy flux and energy density are displayed. In Figure~\ref{fig:EProdRate}, the energy production rates required at the sources to fuel the observed energy flux are shown as a function of energy for the different primary mass groups. While the contribution of nuclei peaks at most to ${\simeq}\: 5 \times 10^{44}\: {\rm erg}\: {\rm Mpc}^{-3}\: {\rm yr}^{-1}\: {\rm dex}^{-1}$, that of protons is increasing up to ${\simeq}\: 10^{45}\: {\rm erg}\: {\rm Mpc}^{-3}\: {\rm yr}^{-1}\: {\rm dex}^{-1}$ when going down in energy. Extrapolations of the results below 0.63\:EeV are however hazardous, as the functional shape used in Equation~\eqref{eqn:qN} may not hold anymore depending on the specifics of the source environments that govern the nuclear cascade. Integrated above 0.63\:EeV, the total energy production rate is found to be $(10.8 \pm 0.4 )\times 10^{44}\: {\rm erg}\: {\rm Mpc}^{-3}\: {\rm yr}^{-1}$. For completeness, the $X_{\rm max}$ distributions used in this work together with the best-fit models obtained within the benchmark scenario explored here are shown in  Appendix~\ref{app:Xmax} (Figure~\ref{fig:XmaxDistrib}).

The reduced deviance is decomposed into three terms, according to Equation~\eqref{eqn:deviance}. As detailed in Appendix~\ref{app:tables}, the spectral sector leads to an acceptable fit ($D_J+D_{J_{\mathrm{p}}}=37.3$ for $N_J+N_{J_{\mathrm{p}}}=24$ points). As in~\cite{PierreAuger:2017pzq}, the $X_{\mathrm{max}}$ sector is more difficult to fit as evidenced by the value $D_{X_{\mathrm{max}}}=199.5$ for $N_{X_{\mathrm{max}}}=109$ points. The large value of the deviance on $X_{\mathrm{max}}$ reflects the difficulty for hadronic-interaction generators to reproduce the data, as illustrated in Appendix~\ref{app:Xmax}. In general, the benchmark scenario fits the data with similar performances to those obtained by considering only data beyond 5\:EeV. Considering the variation of Equation~\eqref{eqn:qN} that consists in substituting $E_{\mathrm{max}}$ for $E_{\mathrm{max}}/2$ yields the results reported in the fourth line of Table~\ref{tab:resultsDeviance_EPOS}. It is interesting that the main changes concern the spectral index of protons, which becomes $\gamma_{\mathrm{p}}\simeq 2.5$. The astrophysical consequences of these results will be discussed in Section~\ref{sec:discussion}, after the sources of systematic uncertainties have been discussed below.

The inferred mass-dependent energy production rates, tracing directly the abundances of elements escaping from the source environments, rely primarily on the $X_{\mathrm{max}}$ distributions. The systematic uncertainties in $X_{\mathrm{max}}$, which are slightly energy-dependent and range from 6 to 9\:g\:cm$^{-2}$~\citep{PierreAuger:2014sui}, are thus expected to impact the results. A reduction of the $X_{\mathrm{max}}$ scale by $1\:\sigma$, which results in an overall heavier composition on Earth, has as main impact a decrease of the proton and helium contributions by respectively ${\simeq}\:15\%$ and ${\simeq}\:35\%$, with an increase of the spectral index $\gamma_A$ of $+0.7$. This configuration leads to a deviance improvement of ${\simeq}\: 16$ units. On the other hand, an increase of the $X_{\mathrm{max}}$ scale by $1\:\sigma$ deteriorates the quality of the fit in an untenable way (more than 100 units of deviance), as the overall lighter composition implied is then in tension with the width of the $X_{\mathrm{max}}$ distributions. By contrast, the inferred parameters and quality of the fits are mildly altered by the systematic uncertainties affecting the energy spectrum, which are dominated by those in the energy scale ($\Delta E/E=14\%$). This is because a change in $E_{\mathrm{max}}^Z$ can be reproduced by slightly different balances regulating the intensity of each nuclear component due to the dependence of the photodissociation threshold with the mass number $A$ of the nuclei.

The interpretation of the $X_{\mathrm{max}}$ distributions heavily relies on the hadronic-interaction generator used in the simulations of extensive air showers. Based on the generator Sibyll2.3, increased (decreased) fractions of intermediate nuclei (protons) are inferred at all energies~\citep{Bellido:2017cgf}. An overall heavier composition is thus implied, in a manner similar to that obtained with the analysis based on EPOS-LHC after a $-1\:\sigma$ shift of the $X_{\mathrm{max}}$ scale is applied. The results reported in Appendix~\ref{app:sibyll2.3} are thus similar to those discussed above when applying the $-1\:\sigma$ shift of the $X_{\mathrm{max}}$ scale.

The range of $E_{\mathrm{max}}^Z$ makes the role of the EBL more important than that of the CMB to control the energy losses by photodissociation of the nuclei. Consequently, the uncertainties affecting both the far-infrared intensity of the EBL and the partial cross sections of channels in which $\alpha$ particles are ejected can alter the results. Considering the EBL model of~\cite{Dominguez:2010bv} instead of~\cite{Gilmore:2011ks} results in an increased far-infrared density of photons, which enhances the intensity of secondary protons from photodissociations \emph{en route}. Most notably, this enhancement is compensated in the fit to the data through an energy production rate of He (CNO) elements that is increased by ${\simeq}\: 25\%$ (${\simeq}\: 45\%$) while that of protons is decreased by ${\simeq}\: 30\%$. Also, the spectral index of protons is increased by $+1$ for a global deviance similar to that of the benchmark scenario. On the other hand, cross sections for photodissociation from~\cite{Puget:1976nz,Stecker:1998ib}, which neglect $\alpha$-particle production, impact mainly the balance of the energy production rate of He (CNO) [Si], changed by ${\simeq}\: +70\%~(-45\%)~[+20\%]$, at the cost of a degraded deviance by 12 units.

%As already stressed, the absence of any firmly-established prescription for modeling the upper end of the acceleration process forces us to use a generic functional shape for $f_{\mathrm{supp}}(E,Z)$. The natural extended width that controls the rate at which the broken-exponential function adopted in Equation~\eqref{eqn:fsupp} falls off has an influence on both $E_{\mathrm{max}}^Z$ and $\gamma$, due to the narrow energy range probed between $5\:$EeV and $E_{\mathrm{max}}^Z$ in the case of nuclei. An alternative functional shape that guarantees a suppression taking place over a much narrower energy range is given by $f_{\mathrm{supp}}(E,Z)=\left[\mathrm{cosh}\left(E/E_{\mathrm{max}}^Z\right)^2\right]^{-1}$. Such a change induces an increase of $E_{\mathrm{max}}^Z$ by a factor ${\simeq}\: 3$ as well as an increase of $\gamma_A$ by $+1.1$. However, the quality of the fit is then degraded by ${\simeq}\: 30$ units of deviance. 

Finally, we studied the impact of the redshift evolution of the UHECR luminosity density, considered as flat on large-spatial scales in the benchmark scenario. Alternatively to a flat evolution, we tested a scenario where the evolution strictly follows the stellar-mass distribution on large spatial scales \citep{Madau:2014bja}. The only notable difference with the benchmark scenario is a decrease by 20\% of the emissivity of all components, which can be understood from the lower average source distance resulting from the decrease of stellar-mass density with redshift at $z>1$. Another attractive scenario is to assume an evolution proportional to the redshift-dependent star-formation rate. Using the corresponding $S(z)$ function derived in~\cite{Madau:2014bja}, the most notable differences are an increase of the overall energy production rate by a factor $2.5$ and a change in the balance of H (He) [CNO] \{Si\} by ${\simeq}\: +40\% (+90\%) [+350\%] \{+25\%\}$. In such a scenario, the emissivity would be dominated by a soft proton component and a hard CNO component, with lower mass nuclei partly generated by photodissociation \textit{en route} from the cosmic-noon epoch at $z>1$. The quality of the fit is almost unchanged for a source evolution following that of stellar mass or star formation compared to that of the benchmark scenario.

\section{Discussion} 
\label{sec:discussion}

The results presented in the previous section direct the interpretation of the origin of protons below the ankle energy and of the hard spectra of nuclei above this energy. In this picture, the component of protons is of extragalactic origin well below the ankle energy and exponentially suppressed above it, while heavier nuclei steadily take over to the highest energies through a rigidity-dependent maximum-energy scenario. That the protons do not get suppressed when going down in energy at the same fast rate as the heavier nuclei is modeled by a softer spectral index for this population of low-charge primaries. Interestingly, such a behavior qualitatively fits with scenarios of in-source interactions in which copious fluxes of neutrons, which are produced while accelerated charged particles interact with the bath of photons permeating the sources, escape freely the electromagnetic fields. 

\subsection{Comparison with other works}

Although this scenario has already received considerable attention in recent years~\citep{Unger:2015laa,Globus:2015xga,Biehl:2017zlw,Zhang:2017moz,Fang:2017zjf,Supanitsky:2018jje,Boncioli:2018lrv,Muzio:2021zud}, our results provide a robust set of constraints for the spectral indices $\gamma_{\rm p}$ and $\gamma_A$ as well as for the energetics of the sources and for the abundances of elements in their environments. The reason for the robustness of the constraints is twofold. First, the choice to consider the proton spectrum only between 0.63~EeV and 5~EeV allows us to reconstruct the extragalactic component as directly and accurately as possible, without any ``interference'' from the modeling of the  composition of the supposedly upper end of the galactic component. The proton flux in this energy range is indeed generally completed by $\textit{ad hoc}$ mass components of galactic origin so as to match the all-particle flux and, at most, the first and second moments of $X_{\rm max}$ or $\ln A$. This leaves some freedom for the modeled proton flux to deviate from the observed one. Second, the deviation from the widely-used uniform distribution of sources in a comoving volume, through the addition of a local overdensity, allows us to match the prediction with the data on a statistical basis characterized by $D/{\rm ndf}\simeq 237/125$, i.e.\ a reduced deviance on the order of that found in \cite{PierreAuger:2016use} for data strictly above the ankle. The study presented here improves substantially the description of the data, especially the mass-composition sector, compared to previous studies across the ankle reporting goodness-of-fit estimators~\citep{Unger:2015laa,Muzio:2021zud} while only qualitative trends are generally reported in others. One notable difference, beyond the abundances of elements adjusted to the data in our case, concerns the much softer value of the spectral index $\gamma_{\rm p}$, which can be observed to be generally close to $\gamma_{\rm p}=2$ (or even harder) in several studies~\citep{Globus:2015xga,Unger:2015laa,Biehl:2017zlw}. The exploratory finding of $\gamma_{\rm p}\simeq 2.5$ in the case of protons produced exclusively by in-source interactions might point towards environments largely enriched in intermediate-mass nuclei compared to protons.

\subsection{Making up the all-particle spectrum below the ankle feature}

We now comment on the ``upper end of the galactic component'' not addressed here but needed to complete the picture of the ankle energy range.
If protons of extragalactic origin contribute to the sub-ankle component, other elements are needed to make up the all-particle spectrum that, beyond its impressive regularity in the energy region between the second knee near 0.1\:EeV and the ankle near 5\:EeV, hides beneath a complex intertwining of different astrophysical phenomena. Keeping in mind the reliance of the interpretations of $X_\text{max}$ data on the validity of the hadronic interaction models, this intertwining is evidenced by the fraction of elements reported in~\cite{Bellido:2017cgf} as a function of energy that can be described broadly as follows. On the one hand, a steep fall-off of the Fe component is observed well below the ankle energy. This is along the lines of the scenario for the bulk of Galactic cosmic rays characterised by a rigidity-dependent maximum acceleration energy, $E_{\rm max,\: Gal.}^Z \simeq 3Z\:$PeV, for particles with charge $Z$ to explain the knee structures. On the other hand, all hadronic-interaction models indicate the presence of CNO nuclei between the second-knee and ankle energies. This extra-component, which, in the overall scenario explored in this paper, contributes to shape the ankle feature, raises questions. Such intermediate-mass elements could be, for example, fuelled by extragalactic sources different from those producing the bulk of UHECRs above the ankle energy~\citep{Aloisio:2013hya}. Or, they could correspond to a second Galactic component, as first suggested in~\cite{Hillas:2005cs}, for example one resulting from explosions of Wolf-Rayet stars \citep{Thoudam:2016syr}.%, which might, for example, be produced from intermittent sources. 
%In fact, little is known about the confinement of cosmic rays in the Galaxy by the magnetic field at these energies. Adopting the widely-used value of the confinement volume of the Galactic disk, $\pi (20\:\text{kpc})^2\times 300\:\text{pc}\simeq 10^{67}$\:cm$^3$, and adopting a confinement time of the order of $10^{4}$ or $10^5$\:yr for CNO nuclei between $10^{17}$ and $10^{18}$\:eV~\citep{1998AstL...24..139Z,Kaapa:2021}, the putative transient events, such as for instance long gamma-ray bursts releasing out ${\simeq}\: 10^{51}$\:erg every million years in the Milky Way~\citep{Pohl:2011ve}, would supply a luminosity density of a few $10^{-30}$\:erg\:s$^{-1}$\:cm$^{-3}$. This would accommodate the energetics of the He and CNO components shown in Figure~\ref{fig}. Again, the study presented in this paper aims at being exploratory only; further investigation of specific scenarios are beyond the scope of this paper. 

Other observables could help in providing additional signatures to the general scenario outlined above. One of them relies on measurements of large-scale anisotropies, in particular those discriminated by mass. No significant variation of the all-particle flux across the sky has been revealed so far below the ankle energy. However, the direction in right ascension of the first harmonics shows an intriguing constancy in adjacent energy intervals towards the Galactic center~\citep{PierreAuger:2020fbi}. This is potentially indicative of a genuine signal~\citep{Edge:1978rr,Deligny:2018blo}. An interesting possibility to explain such a low level of anisotropy could be that one or several mass components are diluted in the extragalactic component of protons, which is expected to be isotropic to a high level due to their interaction lengths comparable to the cosmological horizon. Some estimates of anisotropy show that the most stringent upper limits up-to-date can then be met for intermediate and heavy nuclei of Galactic origin~\citep{Giacinti:2011ww,PierreAuger:2012gro,PierreAuger:2012xju,Pohl:2011ve}. Given the increasing contribution of the protons to the flux at the ankle energy, the all-particle anisotropy may even decrease with energy~\citep{Deligny:2014opa}, thus providing a mechanism to reduce significantly the amplitude of the vector describing the arrival directions of the whole population of cosmic rays -- which is observationally the only one within reach so far. The measurement of mass-discriminated anisotropies in this energy range is challenging; yet the extension to lower energies of directional-$\langle X_{\mathrm{max}}\rangle$ analyses such as presented in \cite{PierreAuger:2021jlg} could provide elements to further decipher the origin of the intermediate-mass elements below the ankle energy.

Future observations will thus provide elements helping to corroborate the importance of in-source interactions for the interpretation of UHECR data, or to call for alternative interpretations. One of these alternatives consists in assuming that two extragalactic components, from two distinct types of sources, overlap from below the ankle energy to the highest energies~\citep{Aloisio:2013hya,Mollerach:2020mhr,Das:2020nvx,PierreAuger:2021mmt}. If one source type emits only protons, the spectral index of the resulting component was found in~\cite{PierreAuger:2021mmt} to be steep ($\simeq 3.30\pm0.05$), consistent with the results reported here, while the corresponding cutoff energy is not constrained at high energies. Interestingly, a signature of this scenario would be the presence of a sub-dominant component of protons forming less than 10\% of the all-particle flux from $\simeq 10~$EeV up to the highest energies. Such a sub-dominant component could be uncovered, depending on its exact level, with the upgraded instrumentation of the Pierre Auger Observatory in the next years~\citep{Castellina:2019irv,PierreAuger:2016qzd}.

\section*{Acknowledgments} This work was made possible by with the support of the Institut Pascal at Universit\'e Paris-Saclay during the Paris-Saclay Astroparticle Symposium 2021, with the support of the P2IO Laboratory of Excellence (program ``Investissements d’avenir'' ANR-11-IDEX-0003-01 Paris-Saclay and ANR-10-LABX-0038), the P2I axis of the Graduate School Physics of Universit\'e Paris-Saclay, as well as IJCLab, CEA, IPhT, APPEC, the IN2P3 master projet UCMN and EuCAPT ANR-11-IDEX-0003-01 Paris-Saclay and ANR-10-LABX-0038). SM, JB, AC \& OD gratefully acknowledge funding from ANR via the grant MultI-messenger probe of Cosmic Ray Origins (MICRO), ANR-20-CE92-0052. The authors would like to thank colleagues from the Pierre Auger Collaboration for fruitful suggestions, in particular J.~M.~Gonzalez, S.~Mollerach and E.~Roulet for pointing out differences in the normalization of the overdensity parametrization in its cumulated and differential forms, S.~Petrera for providing the latest parameterizations of $X_{\rm max}$ distributions using various hadronic interaction models, and R.~Clay and A.~Di~Matteo for their careful reading and suggestions on the text.
 
\newpage

\appendix

\section{$X_{\mathrm max}$ distributions for the benchmarck scenario}
\label{app:Xmax}

\begin{figure}[h]
\centering
\includegraphics[width=\columnwidth]{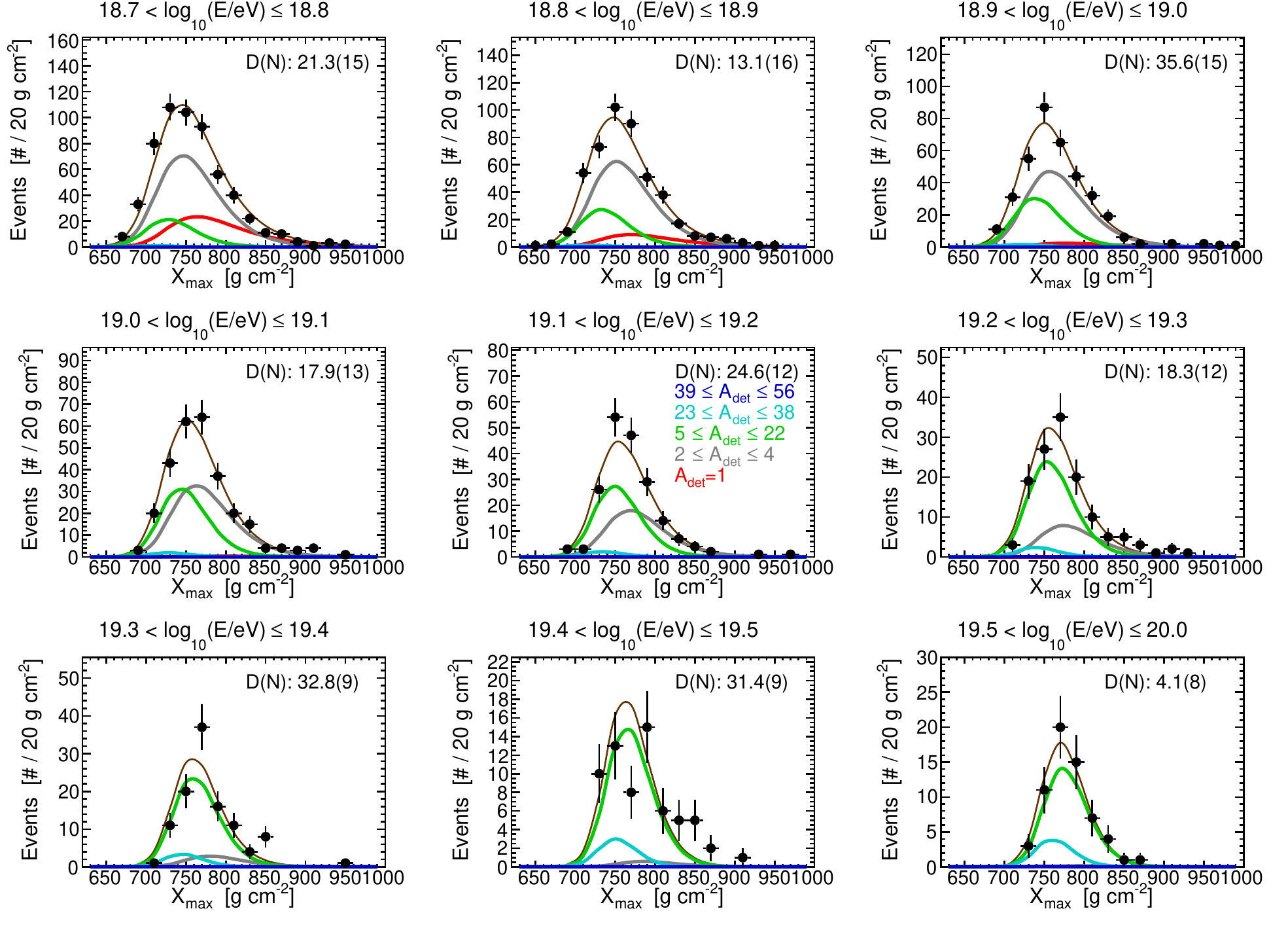}
\caption{Distributions of shower-depth maximum, $X_{\rm max}$, with best-fit models. The contribution of five detected mass groups are displayed with solid colored lines, as labeled in the Figure. The sum of the five contribution is shown in brown. The deviance (D) and number of points (N) is shown in the top-right corner of each plot for each energy bin. Note that one event at 1,070\,g\,cm$^{-2}$ is out of the displayed $X_{\rm max}$ range for $18.9 < \log_{\rm 10}(E/{\rm eV}) \leq 19.0$.}
\label{fig:XmaxDistrib}
\end{figure}

Figure \ref{fig:XmaxDistrib} shows the distributions of shower-depth maximum, $X_{\rm max}$, together with the best-fit models obtained within the benchmark scenario (EPOS-LHC hadronic interaction model).

\clearpage

\section{Deviance for the benchmark scenario}
\label{app:tables}

Table~\ref{tab:resultsDeviance_EPOS} shows the breakdown of deviance for the two scenarios: $\gamma_{\rm p}=\gamma_{A}$ and $\gamma_{\rm p}\neq\gamma_{A}$. The deviance is broken down in three terms as shown in Equation~\eqref{eqn:deviance}.

\begin{deluxetable}{c|ccc|c}[!h]
\tablecaption{Breakdown of deviance for the benchmark scenario. \label{tab:resultsDeviance_EPOS}}
\tablehead{
\colhead{Scenario} 	&	\colhead{$D_J\ (N_J)$} 	&	\colhead{$D_{J_\text{p}}\ (N_{J_\text{p}})$}	&	\colhead{$D_{X_{\rm max}}\ (N_{X_{\rm max}})$} 	&	\colhead{$D/\text{ndf}$}}
\startdata
$\gamma_{\rm p}\neq\gamma_{A}$  &  $36.4\ (15)$ & $  \phantom{00}0.9 \ (9)$ & $199.5\ (109)$ & $236.8/125$ \\
$\gamma_{\rm p}=\gamma_{A}$ &  $ \ 25.9\ (15)$ & $632.0 \ (9)$ & $ 204.8 \ (109)$ & $ 862.7/126 $ \\
\enddata
\tablecomments{Column one and five list the scenarios and deviance as in Table~\ref{tab:epos-results}. Columns 2--4 provide a breakdown of the deviance as in Equation~\eqref{eqn:deviance} and the number of points $N$ associated to each term.}
\end{deluxetable}

\section{Best-fit results for Sibyll2.3c}
\label{app:sibyll2.3}

\begin{figure}[h]
\centering
\includegraphics[width=0.5\columnwidth]{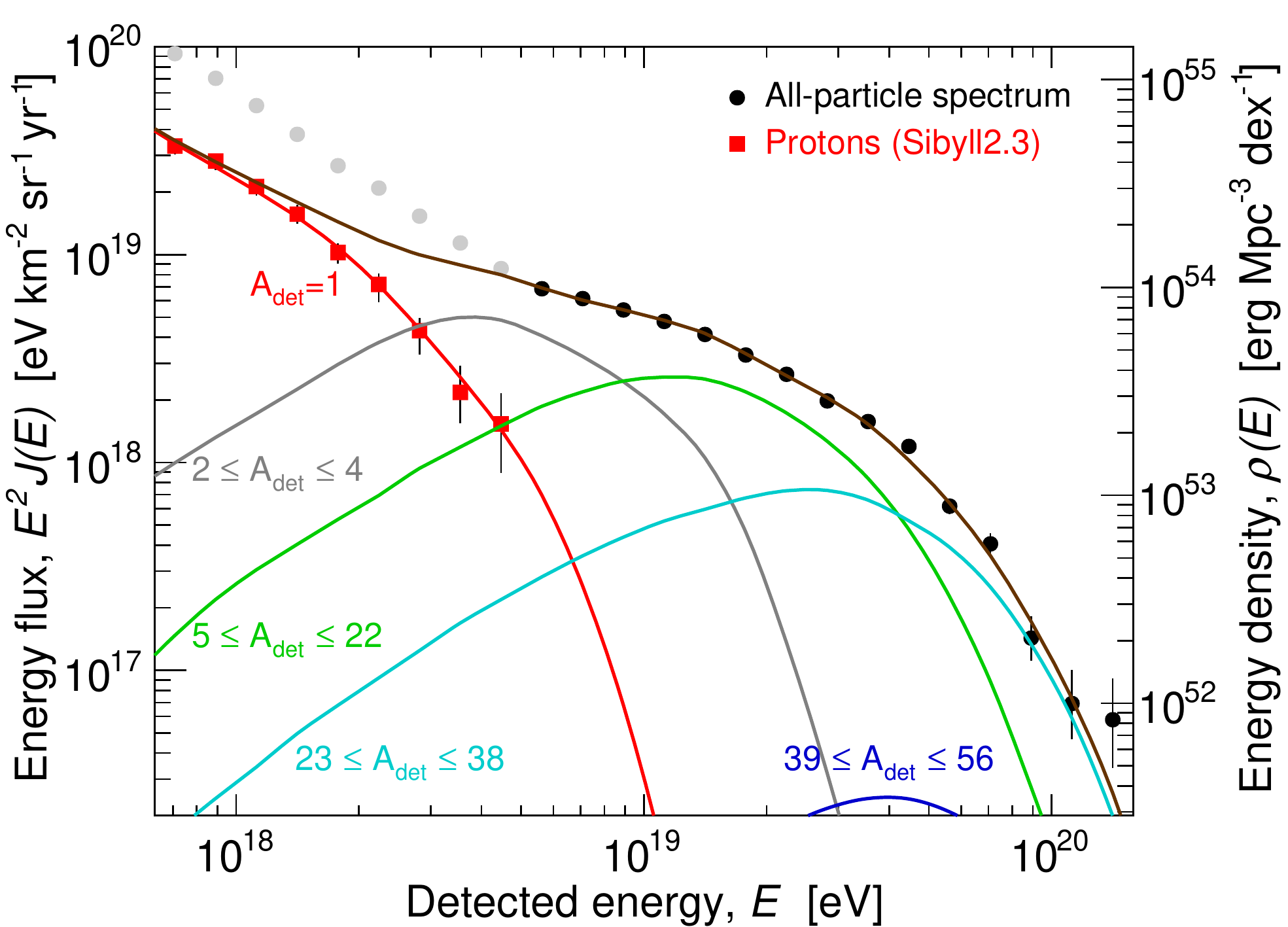}
\caption{Energy flux at Earth as a function of energy, as modeled by the best-fit parameters for a variation around the benchmark scenario using the Sibyll2.3c hadronic interaction model. The all-particle spectrum and proton component are shown as black circles and red squares, respectively. Lighter points are not included in the fit. The best-fit components obtained for five detected mass groups are displayed with solid colored lines, as labeled in the Figure.}
\label{fig:FitFluxSibyll}
\end{figure}

Tables~\ref{tab:sib-results} and~\ref{tab:resultsDeviance_Sibyll} shows the best-fit parameters and breakdown of deviance using Sibyll2.3c as hadronic interaction model instead of EPOS-LHC. As well, Figure~\ref{fig:FitFluxSibyll} shows the energy flux at Earth as a function of energy in the case of Sibyll2.3c. Although the latest version of Sibyll is Sibyll2.3d~\citep{Riehn:2019jet}, Sibyll2.3c is used here as a matter of consistency to match the analysis of \cite{Bellido:2017cgf}, which provides the fractions used to derive the proton flux. For completeness, the analysis has also been run using Sibyll2.3d. It results in similar best-fit parameters, with, however, a composition deviance increased by $\sim 20$ units.

\begin{deluxetable}{c|cccccccc|c}[h]
\tablecaption{Best-fit parameters and minimum deviance for Sibyll2.3c. \label{tab:sib-results}}
\tablehead{
\colhead{Scenario} 	& \colhead{$\gamma_{\rm p}$} 	& \colhead{$\gamma_{A}$} 	&	\colhead{$\log_{10}\: E_{\rm max}\:\text{[eV]}$}	&	\colhead{$\bar{\mathcal{L}}_{\rm p}/\mathcal{L}_{0}$} 	&	\colhead{$\bar{\mathcal{L}}_{\rm He}/\mathcal{L}_{0}$} 	&	\colhead{$\bar{\mathcal{L}}_{\rm N}/\mathcal{L}_{0}$} 	&	\colhead{$\bar{\mathcal{L}}_{\rm Si}/\mathcal{L}_{0}$} 	&	\colhead{$\bar{\mathcal{L}}_{\rm Fe}/\mathcal{L}_{0}$} 	 	&	\colhead{$D/\text{ndf}$}}
\startdata
$\gamma_{\rm p}\neq\gamma_{A}$ & $3.54 \pm 0.25$    	& $ \phantom{-}0.27\pm 0.08$  	&	 $18.36 \pm 0.04$  &	 $2.97 \pm 0.26$	&	 $1.64 \pm 0.11$ 	&	 $2.07\pm 0.18$ 	&	 $1.12 \pm 0.14$  	&	 $0.03 \pm 0.05$  	&	  $216.7/125$  	\\
$\gamma_{\rm p}=\gamma_{A}$ & $\gamma_{A}$ & $ -0.85\pm 0.15$  &	$ 18.15\pm 0.02 $  &	$ 0.55\pm0.10 $	& $ 1.46\pm0.10 $ &	 $ 2.20\pm0.10 $ &	 $ 0.78\pm 0.19$  	& $ 0.13\pm 0.10$  & $369.8/126 $\\
\enddata
\end{deluxetable}

\begin{deluxetable}{c|ccc|c}[h]
\tablecaption{Breakdown of deviance for Sibyll2.3c. \label{tab:resultsDeviance_Sibyll}}
\tablehead{
\colhead{Scenario} 	&	\colhead{$D_J\ (N_J)$} 	&	\colhead{$D_{J_\text{p}}\ (N_{J_\text{p}})$}	&	\colhead{$D_{X_{\rm max}}\ (N_{X_{\rm max}})$} 	&	\colhead{$D/\text{ndf}$}}
\startdata
$\gamma_{\rm p}\neq\gamma_{A}$  &  $19.7  \ (15)$ & $ \phantom{00}1.2 \ (9)$ & $195.8\ (109)$ & $216.7/125$ \\
$\gamma_{\rm p}=\gamma_{A}$ &  $ 18.6 \ (15)$ & $ 142.0 \ (9)$ & $ 209.2 \ (109)$ & $ 369.8/126 $ \\
\enddata
\end{deluxetable}

\newpage
\bibliographystyle{aasjournal.bst}
\bibliography{biblio}

\end{document}